\documentclass[twocolumn, showpacs, superscriptaddress, nofootinbib]{revtex4-1}
\usepackage[utf8]{inputenc}
\usepackage[english]{babel}
\usepackage{graphicx}
\usepackage{float}
\usepackage{amsmath}
\usepackage{mathtools}			
\usepackage{bm}
\usepackage{environ}
\usepackage{comment}
\usepackage{grffile}
\usepackage{xcolor}
\usepackage{booktabs}			
\usepackage{placeins}			
\usepackage{amsfonts}
\usepackage{amssymb}
\usepackage{wasysym}
\usepackage{dsfont}
\usepackage{color,soul}
\usepackage{dcolumn}			
\usepackage{bm}				
\usepackage[pdftex, colorlinks, citecolor=blue, linkcolor=blue, urlcolor=blue]{hyperref}
\usepackage[all]{hypcap}			
\usepackage{paralist}			
\usepackage[version=4]{mhchem}	
\usepackage{tikz-cd}				
\usepackage{chemfig}

\definecolor{myred}{rgb}{0.8,0.1,0.1}
\definecolor{mygre}{rgb}{0.0,0.5,0.2}
\definecolor{myblu}{rgb}{0.0,0.0,0.9}

\graphicspath{{./figures/}}

\begin{document}

\title{Multiple Notch ligands in the synchronization of the segmentation clock}
\author{Marcos Wappner} 
\affiliation{Instituto de Investigaci\'on en Biomedicina de Buenos Aires (IBioBA) – CONICET/Partner Institute of the Max Planck Society, Polo Cient\'{i}fico Tecnol\'ogico, Godoy Cruz 2390, Buenos Aires C1425FQD, Argentina}
%
\author{Koichiro Uriu}
\affiliation{School of Life Science and Technology, Tokyo Institute of Technology, 2-12-1 Ookayama, Meguro, Tokyo 152-8550, Japan}
\affiliation{Graduate School of Natural Science and Technology, Kanazawa University, Kakuma-machi, Kanazawa 920-1192, Japan}
\author{Andrew C. Oates}
\affiliation{Institute of Bioengineering, School of Life Sciences, Swiss Federal Institute of Technology Lausanne (EPFL), Lausanne 1015, Switzerland}
\author{Luis G. Morelli}		
\affiliation{Instituto de Investigaci\'on en Biomedicina de Buenos Aires (IBioBA) – CONICET/Partner Institute of the Max Planck Society, Polo Cient\'{i}fico Tecnol\'ogico, Godoy Cruz 2390, Buenos Aires C1425FQD, Argentina}
%
%

\date{\today}
%
\begin{abstract} 
Notch signaling is a ubiquitous and versatile intercellular signaling system that drives collective behaviors and pattern formation in biological tissues.
During embryonic development, Notch is involved in generation of collective biochemical oscillations that form the vertebrate body segments, and its failure results in embryonic defects.
Notch ligands of the Delta family are key components of this collective rhythm, but it is unclear how different Delta ligands with distinct properties contribute to relaying information among cells.
Motivated by the zebrafish segmentation clock, in this work we propose a theory describing interactions between biochemical oscillators, where Notch receptor is bound by both oscillatory and nonoscillatory Delta ligands. 
Based on previous in vitro binding studies, we first consider Notch activation by Delta dimers.
This hypothesis is consistent with experimental observations in conditions of perturbed Notch signaling. 
Then we test an alternative hypothesis where Delta monomers directly bind and activate Notch, and show that this second model can also describe the experimental observations.
We show that these two hypotheses assign different roles for a non-oscillatory ligand, as a binding partner or as a baseline signal.
Finally, we discuss experiments to distinguish between the two scenarios.
Broadly, this work highlights how a multiplicity of ligands may be harnessed by a signaling system to generate versatile responses.
\end{abstract}
\maketitle

%
Intercellular signals coordinate cellular dynamics across tissues to generate collective behaviors and patterns~\cite{marks, alberts, wolpert}. 
A key signaling pathway is Notch, a versatile system controlling a wide variety of processes~\cite{andersson11, guruharsha12, henrique19, sachan23}.
During embryonic development, Notch signaling coordinates cellular dynamics across tissues to generate patterns~\cite{wolpert}.

\begin{figure}[t]
\centering
\includegraphics[width=\columnwidth]{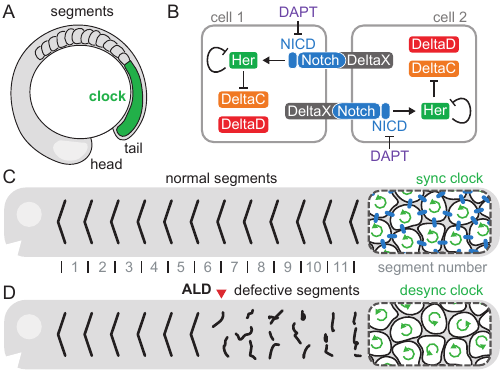}
\caption{
(A) Schematic representation of zebrafish embryonic segmentation showing the unsegmented clock tissue (green) and forming segments. 
(B) Delta-Notch intercellular coupling synchronizes individual oscillators.
(C, D) Schematic zebrafish showing (C) normal segment boundaries (black lines) and (D) defective boundaries (broken black lines). Red arrowhead indicates the onset of defective segments (ALD, see text). 
Magnifications of the tail tip show (C) synchronized individual oscillators (coherent circular arrows) in the presence of coupling (blue bars) and 
(D) desynchronized oscillators (incoherent circular arrows) in its absence.
}
\label{fig:model}
\end{figure}

An attractive model system to study Notch signaling dynamics is the segmentation clock, a collective rhythm that governs the segmentation of the vertebrate body~\cite{oates12, shimojo16, pourquie22, venzin20}, Fig.~\ref{fig:model}A.
%
%
This rhythm arises from a population of cellular genetic oscillators~\cite{jiang00, horikawa06, riedel-kruse07, delaune12, okubo12, isomura17, yoshioka-kobayashi20, yamanaka22, krol11}.
In zebrafish, cell autonomous oscillations of {\it her}/{\it hes} gene products are driven by a delayed negative feedback loop~\cite{webb16, lewis03, monk03, jensen03, schroter12, trofka12}.
%
Additionally, oscillating {\it her} genes transcriptionally regulate the Notch ligand DeltaC, which also displays oscillating gene expression patterns~\cite{jiang00}.
The ligand binds Notch receptors in neighboring cells, resulting in the proteolytic cleavage and release of the Notch intracellular domain (NICD), which translocates to the nucleus and contributes to the activation of \textit{her} genes in these neighboring cells, Fig.~\ref{fig:model}B.
This results in an oscillating signal, transferring the information of oscillation state from one cell to another.

It is thought that in this way, individual oscillators are synchronized by Notch signaling across the unsegmented tissue, Fig.~\ref{fig:model}C.
Zebrafish embryos with mutations in DeltaC or Notch1a, or which have been treated with the small molecule inhibitor DAPT to block Notch receptor cleavage, Fig.~\ref{fig:model}B, form a few normal segments and then start making defects~\cite{vaneeden96, jiang00,holley02,julich05,riedel-kruse07}.
The desynchronization hypothesis postulates that defective segments are caused by a loss of synchrony in the mutants~\cite{jiang00}, Fig.~\ref{fig:model}D. 
The clock starts in synchronized state~\cite{riedel-kruse07, venzin23}, so a few normal segments can be formed before oscillators drift out of synchrony due to gene expression noise in the absence of intercellular coupling.

The mutant of DeltaD, another Notch ligand present in the zebrafish segmentation clock, also has a phenotype consistent with the desynchronization hypothesis~\cite{delaune12}.
However, DeltaD does not display oscillatory patterns and is not known to be regulated by Her proteins~\cite{oates02,hans02,liao16}.
%
%
It is intriguing that such a constant DeltaD signal, without rhythmic temporal information, has a similar effect on synchronization.

It has been shown that DeltaC and DeltaD associate in vitro and colocalize in vivo~\cite{wright11}.
Additionally, DeltaC and DeltaD homodimers also form in vitro, albeit with a much weaker affinity~\cite{wright11}.
Thus, it is possible that DeltaC and DeltaD form dimers in vivo, and it is these dimers that can bind and activate the Notch receptors in neighboring cells. 
With this hypothesis, the role of DeltaD would be to provide a necessary binding partner for DeltaC~\cite{wright11}.

Previous theoretical work addressed diverse aspects of Notch signaling in the segmentation clock.
Phase theories that coarse grain the molecular details of oscillator coupling have been successfully used to study mean field synchronization dynamics~\cite{riedel-kruse07}, the effects of coupling delays on spatiotemporal dynamics~\cite{morelli09, herrgen10, ares12}, the impact of oscillator mobility~\cite{uriu10, uriu13}, perturbations~\cite{murray13}, asymmetric pulsed coupling~\cite{roth23, ho24} and Notch signaling modes~\cite{pfeuty22, jorg18}. 
Other generic theories consider the possibility that the collective rhythm arises from Notch driven excitable oscillators~\cite{hubaud17, ma09, masamizu06}.
Further work included molecular aspects of Notch signaling, for example to demonstrate synchronized biochemical oscillations~\cite{lewis03, uriu09a, ay13, tiedemann14}, to study Notch receptor modulation~\cite{cinquin03}, amplitude death by coupling delays~\cite{yoshioka-kobayashi20}, and the effects of noise and cell divisions~\cite{horikawa06}, to describe kinematic gene expression waves~\cite{uriu09b, ay14}, and to reveal the effects of noisy communication channels~\cite{jorg18}.
While these previous studies shed light on different aspects of Notch signaling in the segmentation clock, they did not consider distinct roles of ligands.
%
%
%

In this work, we formulate a theory of Delta-Notch coupling in the zebrafish segmentation clock based on existing experimental evidence. 
We use a mean field approach to study synchronization under different scenarios to investigate the role of the ligand DeltaD. 
We first explore the hypothesis that binding of Delta dimers to Notch receptors mediates synchronization.
We show the roles of different dimers in accounting for different aspects of experimental data. 
We then formulate a monomer binding theory and show how it may also explain the role of DeltaD.
We finish with two competing hypotheses and discuss their predictions, suggesting experiments that may be able to distinguish among them.

\vspace{0.5 cm}
\noindent
{\bf \normalsize Oscillator and signal reception \\}
%
In the zebrafish segmentation clock, genes of the \textit{her}/\textit{hes} family have binding sites for their own products, that can act as transcriptional repressors~\cite{oates12}.
It has been proposed that genetic oscillations occur as a consequence of delayed inhibition of these \textit{her} genes~\cite{bessho01, oates02, henry02, bessho03, lewis03, monk03, jensen03, schroter12}.
It is thought that a network of \textit{her} family protein dimers constitutes the core oscillator~\cite{schroter12, trofka12}. 
Here we focus on the roles of the coupling network components, so we take a parsimonious approach in the description of this core oscillator.
A single variable $h_i(t)$ stands for the protein concentration of a generic \textit{her} gene in cell $i$, with $i=1,\ldots,N_c$ and $N_c$ the total number of cells.
The core oscillator dynamics is
\begin{equation}
    \label{eq:h}
    \dot h_i = -h_i + \beta_H 
    \underbracket[.02cm][.2cm]{ \frac{1} {1 +  \left[ h_i(t- \tau_i)\right] ^{\eta}} }_{f_-(h_i(t-\tau_i))}
    \underbracket[.02cm][.2cm]{ \frac{1 +\alpha \left[ \sigma s_i( t - \tau_i) \right] ^{\eta}}
            { 1 + \left[ \sigma s_i( t - \tau_i) \right] ^{\eta} } }_{f_+(s_i(t-\tau_i))} \, .
\end{equation}
The first term describes protein degradation, with unit rate since we set the inverse degradation rate of Her proteins as the timescale to render the system dimensionless, see SM.
The second term describes protein synthesis, with dimensionless basal synthesis rate $\beta_H$.
Synthesis is modulated by a regulatory function, with a negative feedback $f_{-}$ resulting from autoinhibition~\cite{monk03, jensen03}, and a positive regulation $f_{+}$ describing the effect of integrated signals $s_i(t)$ from other cells.
Delayed synthesis regulation accounts for the fact that synthesis rate at time $t$ depends on the concentration of regulatory factors at time $t-\tau_i$, where $\tau_i$ is the time it takes to produce the Her protein in cell $i$.
The values of these synthesis delays are sampled from a normal distribution to capture the effect of gene expression noise~\cite{morelli07}, see SM.
Since the period of oscillations is determined to first order by the synthesis delay~\cite{lewis03, morelli07, novak08}, with this variability in the delays we introduce a variability in the period of individual oscillators.
The intensity of the activation $\alpha$ is the fold change in synthesis rate caused by signals $s_i$.
The dimensionless concentration scale $\sigma$ sets a threshold for signal driven synthesis activation.
The analogous concentration scale for the binding of Her proteins in $f_{-}$ is used as a concentration scale for the variables.
The Hill exponent $\eta$ is an effective nonlinearity accounting for intermediate molecular interactions.
The particular form of the regulatory function in Eq.~(\ref{eq:h}) results from assuming that binding of Her proteins and the NICD to the promoter of the \textit{her} gene occurs independently, see SM.

Communication between cells occurs through the binding of ligands from other cells to Notch receptors, eliciting the NICD signal in the receiving cell.
The dynamics of signal $s_i$ and Notch receptor concentration $n_i$ are given by 
\begin{equation} \label{eq:s}
\dot s_i = -\delta_S s_i + K n_i,
\end{equation}
\begin{equation} \label{eq:n}
\dot n_i = -\delta_N n_i + \beta_N - K n_i,
\end{equation}
where $\delta_S$ and $\delta_N$ are the dimensionless decay rates of the signal and receptor respectively and $\beta_N$ is the Notch synthesis rate. 
The second term in Eq.~(\ref{eq:s}) describes signal production due to the binding of ligands to Notch receptors, with $K$ a total ligand binding activity.
%
%
Next, we specify contributions of different Notch ligands to $K$.

\begin{figure}[t!]
\centering
\includegraphics[width=\columnwidth]{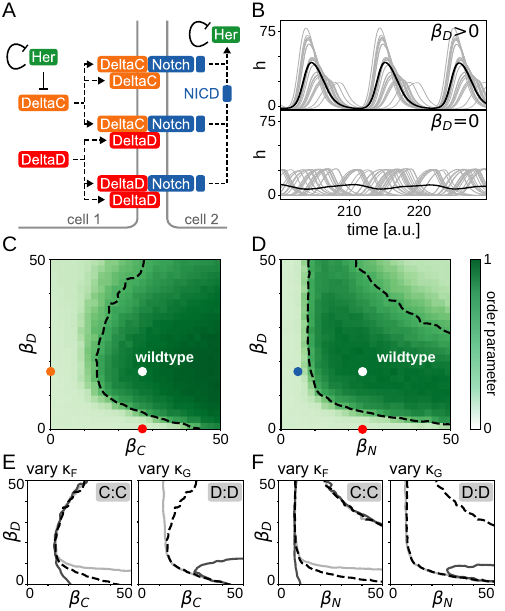}
\caption{Dimer binding hypothesis is compatible with desynchronization phenotypes in steady state. 
(A) All dimers from DeltaC and DeltaD form and can bind and activate Notch. 
(B) Steady state individual oscillations $h_i(t)$ (grey lines, 25 out of 100 are displayed) and mean field $\bar h(t)$ (black line) for $\beta_D > 0$ (top) and $\beta_D = 0$ (bottom).
(C, D) Steady state order parameter $\bar R$ in terms of 
(C) DeltaC and DeltaD synthesis rates, and 
(D) Notch and DeltaD synthesis rates. Dashed line indicates the boundary of the sync region, defined by $\bar R = R_T = 0.6$. 
Dots indicate wildtype parameters (white), and mutants for DeltaC (orange), DeltaD (red), and Notch (blue).
Parameter values for these conditions are determined below from desynchronization dynamics in Fig.~\ref{fig:desync}.
(E, F) Changes to the sync region boundary in terms of 
(E) DeltaC and DeltaD synthesis rates, and 
(F) Notch and DeltaD synthesis rates, 
caused by varying Notch binding rates.
Left: varying C:C Notch binding rate, $\kappa_F = $0.002 (light), 0.02 (dashed black), 0.2 (dark).
Right: varying D:D Notch binding rate, $\kappa_G = $0.002 (light), 0.009 (dashed black), 0.2 (dark).
Dashed black lines are the same as in (C, D).
(C-F) are $25 \times 25$ pixels, averaged over 10 independent realizations. 
Wildtype parameter values in SM.
}
\label{fig:dimers}
\end{figure}

\vspace{0.5 cm}
\noindent
{\bf \normalsize Notch signaling activated by dimers \\}
\label{sec:homodimers}
Following the proposal that DeltaC and DeltaD form dimers~\cite{wright11}, Fig.~\ref{fig:dimers}A, the monomer concentrations $c_i(t)$ and $d_i(t)$ follow
%
%
\begin{align}
    \dot c_i &= - \delta_C c_i + \beta_C \frac{1}{1 + \left[ \gamma h_i (t - \tau_C) \right]^{\eta_c} } \nonumber
        \\ &\qquad \qquad \qquad
        + \lambda_E^- e_i - \lambda_E^+ c_i d_i + \lambda_F^- f_i - \lambda_F^+ c_i^2 \, ,
       \label{eq:homodimer.c} \\
    \dot d_i &= - \delta_D d_i + \beta_D 
        + \lambda_E^- e_i - \lambda_E^+ c_i d_i + \lambda_G^- g_i - \lambda_G^+ d_i^2 \, ,
        \label{eq:homodimer.d} 
\end{align}
including terms describing degradation with rates $\delta_C$ and $\delta_D$, and terms describing synthesis with basal rates $\beta_C$ and $\beta_D$.
We include a modulation of the synthesis rate of DeltaC by the oscillator, since DeltaC promoter contains binding sites for \textit{her} products~\cite{schroter12} and mRNA patterns show signatures of oscillatory expression~\cite{oates02}. 
The concentration of Her in this modulation is evaluated at the past time $t-\tau_C$ to account for the time required to produce DeltaC molecules.
The dimensionless concentration scale $\gamma$ is the threshold for the onset of \textit{deltaC} promoter repression by Her, and Hill exponent $\eta_c$ is the effective nonlinearity of this repression.
DeltaD is not known to be transcriptionally regulated by Her during somitogenesis and its mRNA patterns do not display signatures of oscillation~\cite{oates02}, so we do not include a modulation of DeltaD synthesis. 
Additional terms describe dimerization, where variables $e_i(t)$, $f_i(t)$ and $g_i(t)$ describe the concentrations of DeltaC:DeltaD, DeltaC:DeltaC and DeltaD:DeltaD dimers respectively, and $\lambda_E^{\pm}$, $\lambda_F^{\pm}$ and $\lambda_G^{\pm}$ are association ($+$) and dissociation ($-$) rates.
Dimer dynamics are
\begin{align}
    \dot e_i &= -\delta_E e_i + \lambda_E^+ c_i d_i - \lambda_E^- e_i - \kappa_E e_i \bar n     \, ,
	\label{eq:heterodimer.e} \\
    \dot f_i &= - \delta_F f_i - \lambda_F^- f_i + \lambda_F^+ c_i^2 - \kappa_F f_i \bar n   \, ,
	\label{eq:homodimer.f}  \\
    \dot g_i &= - \delta_G g_i - \lambda_G^- g_i + \lambda_G^+ d_i ^2 - \kappa_G g_i \bar n  \, ,
	\label{eq:homodimer.g} 
\end{align}
resulting in total ligand binding activity
\begin{equation}    
    \label{eq:homodimer.K} 
    K = \kappa_E \bar e + \kappa_F \bar f + \kappa_G \bar g \, ,
\end{equation}
where $\delta_E$, $\delta_F$ and $\delta_G$ are dimer degradation rates, and $\kappa_E$, $\kappa_F$ and $\kappa_G$ are the binding rates of dimers to Notch receptors.
%
%
The bar notation $\bar{x}$ represents the average concentration of species $x$ over all cells that interact with cell $i$.
This average results from the assumption that the cell surface available for signaling is shared equally among all neighbors, see SM.

In this work we describe synchronization dynamics in the posterior tip of the elongating body axis, the so-called tailbud, which is a relatively homogeneous region of the tissue oscillating in synchrony~\cite{soroldoni14}.
Cell mixing is prevalent within this region~\cite{mara07, uriu17, fulton22}, and theory has shown that mixing causes an effective mean field regime~\cite{uriu13, petrungaro19} that enhances synchronization~\cite{uriu10, uriu17}.
Thus, in this work we consider a mean field description of the tail, where all $N_c$ cells interact with every other cell, so
\begin{equation}    \label{eq:mf}
    \bar x = \frac{1}{N_c} \sum_{i=1}^{N_c} x_i  \, .
\end{equation}
%
%
%
%

Together with Eqs.~(\ref{eq:h}-\ref{eq:n}), Eqs.~(\ref{eq:homodimer.c}-\ref{eq:homodimer.K}) and the mean field assumption~(\ref{eq:mf}) complete the model for this multiple dimer binding scenario. 
All equations, variables and parameters are dimensionless, see SM.
In this model, negative feedback-induced autonomous oscillation of Her directly drives the oscillation of DeltaC, Eqs.~(\ref{eq:h}) and (\ref{eq:homodimer.c}).
This in turn may translate into an oscillation of some dimers, Eqs.~(\ref{eq:heterodimer.e}) and (\ref{eq:homodimer.f}), causing the ligand binding activity $K$ to oscillate, Eq.~(\ref{eq:homodimer.K}).
Oscillatory binding activity drives an oscillatory signal in contacting cells, Eqs.~(\ref{eq:n}, \ref{eq:s}), which ends up modulating the Her regulatory function, Eq.~(\ref{eq:h}).
In this way, the oscillatory state of one cell is communicated to other cells.

\vspace{0.5 cm}
\noindent
{\bf \normalsize Steady state synchronization maps \\}
We set parameter values following constraints from experimental observations and theoretical considerations which result in synchronized oscillations, Fig~\ref{fig:dimers}B (top), SM and Fig.~S1A.
Next, we seek to describe loss of synchrony as observed in phenotypes of mutant conditions~\cite{jiang00}.
Mutant conditions of signaling components, as well as other perturbations, can be described in the model by altering the corresponding synthesis rates $\beta_C$, $\beta_D$ and $\beta_N$.
For instance, the DeltaD mutant could be described by setting $\beta_D=0$.
We find that synchronization is lost in this condition, matching the experimental evidence for desynchronization~\cite{delaune12}, even though DeltaD is not regulated by the oscillator, Fig.~\ref{fig:dimers}B (bottom).
To quantify how different perturbations may affect the collective rhythm, we introduce a synchronization index, the Kuramoto order parameter $R$ (Methods)~\cite{strogatz}, taking values between 0 (desynchronized) and 1 (synchronized). 
We construct steady state synchronization maps that show the stationary value of this order parameter after transients have elapsed, Fig.~\ref{fig:dimers}C, D.
In these maps, we can introduce synchronization boundaries that separate synchronized from desynchronized states by setting a threshold $R_T$ on the value of the order parameter. 
Dashed lines in Fig.~\ref{fig:dimers}C, D indicate this boundary with the choice $R_T = 0.6$.
The presence of segmentation defects in mutant phenotypes for the Delta ligands and the Notch receptor are interpreted as desynchronized states in these maps.
We can see that it is possible to choose a wildtype parameter set within the synchronized region, such that reducing the synthesis rates of the coupling components, will result in drastic loss of steady state synchrony, see dots in Fig.~\ref{fig:dimers}C, D.
In such parametrization of the wildtype, all coupling components are required for steady state synchronization, in agreement with mutant phenotypes for the ligands and the receptor~\cite{delaune12}.
In particular, this result is consistent with a role for DeltaD in the synchronization of the segmentation clock.
Furthermore, we observe that starting from the $\beta_D=0$ condition and increasing either $\beta_C$ or $\beta_N$ we can move back into the synchronized region, see red dot in Fig.~S2A, B.
This predicts a rescue of DeltaD mutant defects with adequate DeltaC or Notch overexpression.

These maps also predict a non-monotonic behavior of the order parameter with DeltaD synthesis rate $\beta_D$, Fig.~\ref{fig:dimers}C, D. 
Extended synchronization maps show that for large $\beta_D$ synchrony is lost, suggesting that non-oscillating components become so abundant that they titrate out the rhythmic DeltaC signal, Fig.~S2A, B.
This is consistent with experiments, where DeltaD overexpression by injection of high mRNA levels caused defective segments~\cite{dornseifer97}, likely due to loss of synchrony.
In contrast, transgenic zebrafish expressing 50 times the level of wildtype DeltaD, driven by the endogenous regulatory regions, did not cause desynchronization phenotypes~\cite{liao16}.
Taken together, these data suggest that the wildtype is placed in a region of the synchronization map with some place for larger DeltaD concentrations before synchrony is lost.
In the theory, DeltaD overexpression can be represented by large $\beta_D$ values.
To test whether the model can account for DeltaD overexpression data, we look for parameters that control the shape of the synchronization region for large values of $\beta_D$.
Boundaries of steady state synchronization regions can be tuned by parameters that control aspects of dimer kinetics, Fig.~\ref{fig:dimers}E, F and Fig.~S3.
For example, increasing the value of $\kappa_F$ changes the point on the $\beta_D = 0$ axis where increasing the values of $\beta_C$ or $\beta_N$ leads to a synchronization transition, left panels in Fig.~\ref{fig:dimers}E, F.
In contrast, the boundary for large $\beta_D$ can be shaped by the value of $\kappa_G$, without significantly changing the synchronization transition at $\beta_D = 0$.
Thus, DeltaD overexpression can be captured in the model while preserving the behavior of the DeltaD mutant condition. 
\begin{figure}[t!]
\centering
\includegraphics[width=\columnwidth]{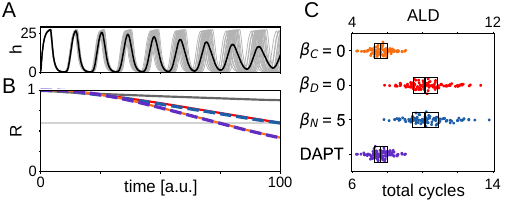}
\caption{Dimer binding hypothesis is compatible with transient desynchronization phenotypes.
%
(A) Individual oscillators $h_i(t)$ (grey lines, 20 out of 100 are displayed) and mean field $\bar h(t)$ (black line) for $\beta_C = 0$.
(B) Order parameter as a function of time from initially synchronized state for different conditions: 
wildtype parameters (dark grey line), 
$\beta_C=0$ (orange line), 
$\beta_D=0$ (red line), 
$\beta_N=5$ (dashed blue line), and 
DAPT treatment (dashed purple line).
Horizontal light grey line marks the threshold value $R_T=0.6$ of the order parameter.
(C) Onset of defective segments for different conditions as indicated. 
Top axis is ALD in cycles from the onset of segmentation, bottom axis is cycles from the onset of oscillations (Methods).
Dots are 100 individual realizations with $100$ cells.
Bars are medians and boxes display the interquartile range.
}
\label{fig:desync}
\end{figure}

%
%
\vspace{0.5 cm}
\noindent
{\bf \normalsize Desynchronization dynamics \\}
In the embryo, the segmentation clock is initiated in a synchronized state~\cite{riedel-kruse07, venzin23}. 
It is thought that in Notch signaling mutants, impaired coupling leads to desynchronization of the collective rhythm and eventually causes the observed defective segments~\cite{jiang00, horikawa06, riedel-kruse07, delaune12}.
Although all mutants of coupling components end up desynchronized, some start making defects earlier than others~\cite{oates05}.
These differences in the onset of defects is often quantified as the anterior limit of defects (ALD)~\cite{oates02}.
While the \textit{deltaC} mutant has a mean ALD of about $5.5$~\cite{vaneeden96, julich05}, \textit{deltaD} and \textit{notch1a} have a mean ALD of $8$~\cite{vaneeden96, riedel-kruse07, liao16}.
There are multiple \textit{notch} genes expressed in the segmentation clock~\cite{westin97}, but segmentation phenotypes have not been reported for \textit{notch} mutants other than \textit{notch1a}.
%
%
This may reflect that the individual contributions from each of the other Notch receptors to synchronization is small. 
This interpretation is consistent with the identical ALD of the \textit{deltaC} mutant and the DAPT assay~\cite{riedel-kruse07}, which should block the cleavage of all Notch receptors.

The differences in ALD can be interpreted as a consequence of different desynchronization rates in the mutant backgrounds.
To test this aspect of mutant phenotypes, we introduce a computational desynchronization assay.
Experimental observations show that \textit{her7} oscillations start synchronized and make a few cycles before the initiation of somitogenesis~\cite{riedel-kruse07}.
Similarly, a Her1 protein reporter shows several synchronized cycles at the segmentation clock onset~\cite{venzin23}.
With this motivation, we set the initial values of all variables to zero for all cells.
Simulations show that oscillations start synchronized, and synchrony decays in the absence of key coupling components, Fig.~\ref{fig:desync}A.
We then look for parameter values $\beta_C$, $\beta_D$ and $\beta_N$ that capture experimental observations of ALD in mutant backgrounds, 
while keeping the rest of parameters as in Fig.~\ref{fig:dimers} (Methods). 
In the theory, the onset of defects can be associated with the threshold value $R_T$ for the Kuramoto order parameter.
We interpret that below this value, synchrony is not enough to produce a normal segment boundary.
We obtain a parameter set for which the desynchronization assay is consistent with ALD measurements of different conditions: the wildtype stays synchronized, threshold crossing occurs soonest for the DeltaC mutant condition, and at later similar times for the DeltaD and Notch mutant conditions, Fig.~\ref{fig:desync}B.
Synchrony decay for DeltaC mutant occurs as fast as in a simulation of saturated DAPT treatment (Methods)~\cite{horikawa06, riedel-kruse07}.
Scaling the time by the cycle duration we can construct ALD plots (Methods), which show good agreement with experimental observations, Fig.~\ref{fig:desync}C.

\begin{figure}[t]
\centering
\includegraphics[width=\columnwidth]{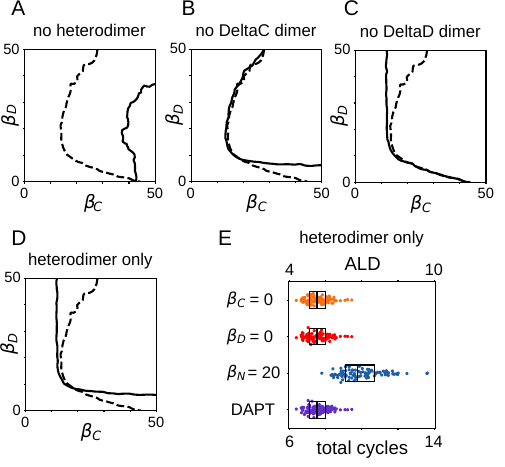}
\caption{Individual dimers influence distinct regions of the synchronization map.
(A-D) Changes to sync region boundary in Fig.~\ref{fig:dimers}C (dashed line) for 
(A) $\lambda_E^{\pm}=0$ and $\kappa_E=0$ (solid line), 
(B) $\lambda_F^{\pm}=0$ and $\kappa_F=0$ (solid line), 
(C) $\lambda_G^{\pm}=0$ and $\kappa_G=0$ (solid line),
(D) $\lambda_F^{\pm}=\lambda_G^{\pm}=0$ and $\kappa_F=\kappa_G=0$ (solid line).
(E) Onset of defective segments for different conditions as indicated. 
Axes, dots and boxes as in Fig.~\ref{fig:desync}C.
Dashed line, number of realizations and other parameters as in Fig.~\ref{fig:dimers}E, F.
}
\label{fig:het}
\end{figure}

\vspace{0.5 cm}
\noindent
{\bf \normalsize The role of different dimers \\}
%
Next, we ask how different dimers contribute to shaping the synchronization region, by alternatively setting to zero their association and dissociation rates together with their Notch binding rates.
%
%
The DeltaC:DeltaD heterodimer is important to generate a wide synchronization region, with a broad range of $\beta_C$ values where the DeltaD mutant would result in loss of synchronization, Fig.~\ref{fig:het}A.
%
%
The DeltaC:DeltaC homodimer further enables synchronization in the low DeltaD region, such that large $\beta_C$ could rescue the DeltaD mutant, Fig.~\ref{fig:het}B.
%
%
In contrast, the DeltaD:DeltaD homodimer can inhibit synchronization for large DeltaD concentrations, Fig.~\ref{fig:het}C.
Excluding both homodimers results in a symmetric synchronization map, in which loss of any ligand is equivalent, Fig.~\ref{fig:het}D.
The synchronization map of this heterodimer only scenario is consistent with the steady state properties of mutants, that is, loss of synchrony.
However, the numerical desynchronization assay displays the same ALD for DeltaC and DeltaD mutants, failing to recapitulate distinct embryonic mutant phenotypes, Fig.~\ref{fig:het}E.

In summary, the hypothesis that DeltaC and DeltaD dimers bind and activate the Notch receptor seems to be consistent with key experimental observations of mutant phenotypes, Figs.~\ref{fig:dimers} and~\ref{fig:desync}.
In this hypothesis, the role of DeltaD is to provide a binding partner for DeltaC.
So, even though DeltaD is not transcriptionally regulated, through its interaction with DeltaC it cooperates in generating an oscillatory signal.
Thus, the heterodimer DeltaC:DeltaD seems to be the key player in this scenario, Fig.~\ref{fig:het}A.
Still, homodimers are required as well to allow for the observed differences in the desynchronization assays of DeltaC and DeltaD mutants, Figs.~\ref{fig:desync}C and~\ref{fig:het}E.

\begin{figure}[t!]
\centering
\includegraphics[width=\columnwidth]{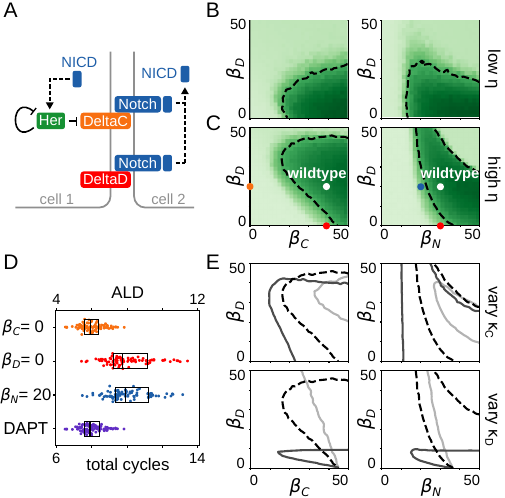}
\caption{Monomer binding  hypothesis is compatible with experimental data. 
(A) DeltaC and DeltaD monomers can bind and activate Notch. 
(B, C) Steady state order parameter $\bar R$ for 
(B) effective nonlinearity $\eta=2.5$ and 
(C) $\eta=7$, in terms of 
(left) DeltaC and DeltaD synthesis rates, and 
(right) Notch and DeltaD synthesis rates. 
Dashed line, dots, color scale and number of realizations as in Fig.~\ref{fig:dimers}C, D.
Wildtype parameter values in SM.
(D) Onset of defective segments for different conditions as indicated. 
Axes, dots and boxes as in Fig.~\ref{fig:desync}C.
(E) Changes to the sync region boundary in terms of 
(left) DeltaC and DeltaD synthesis rates, and 
(right) Notch and DeltaD synthesis rates, 
caused by varying Notch binding rates. 
Top: $\kappa_C = $0.01 (light), 0.03 (dashed black), 0.3 (dark).
Bottom: $\kappa_D = $0.002 (light), 0.01 (dashed black), 0.3 (dark).
Dashed black lines are the same as in (C).
Number of realizations as in Fig.~\ref{fig:dimers}E, F.
}
\label{fig:monomer}
\end{figure}

\vspace{0.5 cm}
\noindent
{\bf \normalsize Notch signaling activated by monomers \\}
In the most common view of Notch signaling, ligands are pictured to bind receptors as monomers~\cite{artavanis99, kopan09}. 
Activation of the Notch receptor by Delta dimers would be a paradigm shift in this view~\cite{wright11, chen23}.
Thus, we wanted to test whether dimerization is necessary to explain the data, or whether monomer binding of the Notch receptor is able to reproduce experimental observations as well.
We formulated an alternative theory, in which single DeltaC and DeltaD proteins can bind Notch receptors as monomers, eliciting a signal in the receiving cell, Fig.~\ref{fig:monomer}A.
Without dimerization, the dynamics of DeltaC and DeltaD concentrations are 
\begin{align}
    \label{eq:monomer.c}
    \dot c_i &= -\delta_C c_i + \beta_C \frac{1}{1 + \left[ \gamma \, h_i(t - \tau_C) \right] ^{\eta} } - \kappa_C c_i \bar{n} \, , \\
    \label{eq:monomer.d}
    \dot d_i &= -\delta_D d_i + \beta_D - \kappa_D d_i   \bar{n} \,.
\end{align}
The degradation and synthesis terms are the same as in previous Eqs.~(\ref{eq:homodimer.c}) and (\ref{eq:homodimer.d}), but instead of dimer association and dissociation terms we now include loss terms due to direct binding of the monomer ligands to Notch receptors, with binding rates $\kappa_C$ and $\kappa_D$.
The resulting ligand binding activity $K(t)$ now has contributions from DeltaC and DeltaD ligands,
\begin{equation}
    \label{eq:monomer.K}
    K = \kappa_C \bar{c} + \kappa_D  \bar{d} \, ,
\end{equation}
where $\bar c$ and $ \bar d$ are DeltaC and DeltaD mean fields, Eq.~(\ref{eq:mf}). 
Together, Eqs.~(\ref{eq:h}-\ref{eq:n}, \ref{eq:monomer.c}-\ref{eq:monomer.K}) comprise the model for the scenario where DeltaC and DeltaD ligands bind and activate the Notch receptor.

Synchronized oscillations are also possible in this monomer binding scenario, Fig.~S1B.
We performed a search in parameter space to establish whether this scenario could be compatible with experimental data.
We found that with a parametrization similar to the dimer scenario, synchronization maps do not have a prominent region where synchronization is lost by removing DeltaD, Fig.~\ref{fig:monomer}B.
However, we found that increasing the effective nonlinearity $\eta$ in the regulatory function, steady state synchronization maps are consistent with the phenotypes of DeltaC, DeltaD and Notch mutants~\cite{jiang00, delaune12}, Fig.~\ref{fig:monomer}C and Fig.~S4.
In particular, there is a broad region in the synchronization map where loss of DeltaD causes desynchronization, Fig.~\ref{fig:monomer}C and Fig.~S2C, D.
Furthermore, the desynchronization assay shows that this hypothesis is also compatible with the different desynchronization rates observed in mutants, Fig.~\ref{fig:monomer}D. 

For small values of $\beta_D$ where synchronization is lost, increasing the values of $\beta_C$ and $\beta_N$ can bring the system to a synchronization transition and synchrony recovery.
These maps also reveal that for large values of $\beta_D$ synchronization is lost, Fig.~\ref{fig:monomer}B,~C.
%
However, the boundary of synchronization regions is controlled by the Notch binding rates $\kappa_C$ and $\kappa_D$, Fig.~\ref{fig:monomer}E.
Decreasing $\kappa_D$ extends the synchronization region towards larger $\beta_D$ values, so synchrony may be preserved under strong DeltaD overexpression~\cite{liao16}.
Increasing $\kappa_C$ generally shifts the synchronization region towards smaller values of the syntheses rates, so strong binding of DeltaC to Notch may reduce the cost of sustaining synchrony in terms of molecular turnover.

\begin{figure}[t]
\centering
\includegraphics[width=\columnwidth]{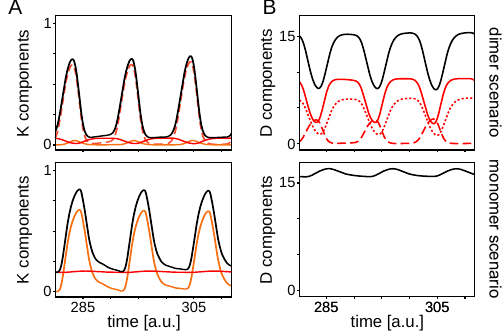}
\caption{Comparison of dimer and monomer scenarios. 
(A) Decomposition of contributions to the total ligand binding activity $K(t)$ (solid black line). 
Top: dimers binding activity $\kappa_Ff(t)$ (orange line), $\kappa_Gg(t)$ (solid red line), $\kappa_Ee(t)$ (dashed red line). 
Bottom: monomers binding activity $\kappa_Cc(t)$ (orange line), $\kappa_Dd(t)$ (red line).
(B) Decomposition of contributions to the total DeltaD dimensionless concentration $d_T(t)$ (solid black line). 
Top: $d(t)$ (solid red line), $e(t)$ (dashed red line), $g(t)$ (dotted red line) in the dimer binding scenario.
Bottom: $d(t) = d_T(t)$ in the monomer binding scenario.
}
\label{fig:totals}
\end{figure}

\vspace{0.5 cm}
\noindent
{\bf \normalsize DeltaD role and dynamics \\}
Going back to our initial motivation, how does an unregulated DeltaD signal, a priori without rhythmic temporal information, contribute to synchronization?
To address this question, we consider the contributions to the ligand binding activity $K(t)$, Fig.~\ref{fig:totals}A.
In the case of dimers binding Notch, large amplitude $K(t)$ oscillations are mainly driven by heterodimer $\kappa_E e(t)$ oscillations.
The homodimer contributions $\kappa_F f(t)$ and $\kappa_G g(t)$ oscillate with a much smaller amplitude since $\kappa_F$ and $\kappa_G$ are an order of magnitude smaller than $\kappa_E$.
Thus, this confirms the role of DeltaD as a binding partner in the heterodimer that is the main activator of Notch receptors.
%
%
In the case of monomers binding Notch, $\kappa_D d(t)$ is relatively constant and $\kappa_C c(t)$ contributes the oscillatory component of $K(t)$.
This suggests that the role of DeltaD in this case is to provide a signaling baseline that shifts $K(t)$ oscillations to larger absolute levels.
In this interpretation, levels of $\kappa_C c(t)$ alone are not enough to synchronize the system, causing the DeltaD mutant phenotype.

These distinct roles of DeltaD in each scenario prompt us to look for possible differences in DeltaD dynamics.
In the dimer binding scenario, DeltaD is present both as a monomer and as component of dimers.
We find that the total DeltaD dimensionless concentration $d_T(t) = d(t) + e(t) + g(t)$ displays pronounced oscillations, Fig.~\ref{fig:totals}B. 
In contrast, in the monomer scenario $d_T(t) = d(t)$ displays small amplitude oscillations.
We quantify the relative amplitude as the peak to trough difference over the average value of the oscillation, which results in a 63\% and 7\% relative amplitude in the dimer and monomer binding scenarios, respectively.

\begin{figure}[t]
\centering
\includegraphics[width=\columnwidth]{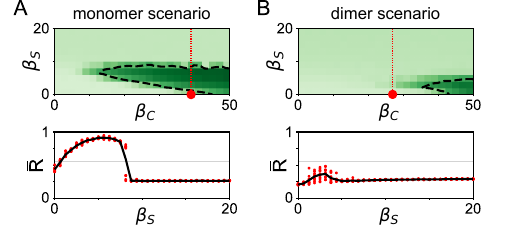}
\caption{NICD overexpression assay in DeltaD mutant predicts distinct outcomes for monomer and dimer scenarios.
Top: Steady state order parameter $\bar R$ in terms of DeltaC and NICD synthesis rates for the (A) monomer and (B) dimer binding scenarios.
Red dot is the DeltaD mutant condition for $\beta_S = 0$ and vertical red dotted line indicates the cut plotted in bottom panels.
Dashed line, color scale and number of realizations as in Fig.~\ref{fig:dimers}C, D. 
Bottom: Steady state order parameter $\bar R$ as a function of $\beta_S$ along the red dotted line in top panels for (A) monomer and (B) dimer binding scenarios.
Red dots are 10 independent realizations and black line is the average.
Horizontal gray line indicates the threshold $R_T$.
Parameters as in corresponding scenario, except for $\beta_D=0$, see SM.
}
\label{fig:rescue}
\end{figure}

Given the distinct role of DeltaD in each scenario, we sought to identify perturbations that could supplant DeltaD in one role but not the other, and might be experimentally realized.
One such possibility is to supply a surrogate baseline for the DeltaC monomer in the absence of DeltaD, but that would not work as the strongly binding partner for DeltaC required in the dimer scenario.
To test this idea, we included in the model an additional synthesis term for $s(t)$, 
\begin{equation}
    \dot s_i = -\delta_S s_i + K n_i + \beta_S  \, ,
\end{equation}
where $\beta_S$ represents an exogenous, unregulated expression of the NICD.
Synchronization maps reveal that there is a range of values of $\beta_S$ that rescue the DeltaD mutant phenotype in the monomer scenario, red dot and dashed line in Fig.~\ref{fig:rescue}A.
In contrast, this NICD overexpression assay does not rescue the mutant in the dimer scenario, Fig.~\ref{fig:rescue}B.
The order parameter along the $\beta_S$ direction shows a range of consistently high values above threshold $R_T$ for the monomer case, while for dimers we observe large fluctuations that stay below threshold, Fig.~\ref{fig:rescue} (bottom).
Another possible perturbation is expressing exogenous unregulated DeltaC ligands, which results in a similar prediction distinguishing the two scenarios, Fig.~S5.
Taken together, these results provide a set of possible experiments that could test the two scenarios.

\vspace{0.5 cm}
\noindent
{\bf \normalsize Discussion \\}
%
%
%
In this work we address the role of DeltaD in the synchronization of the zebrafish segmentation clock.
Motivated by available experimental data, we first consider a model where ligands have to dimerize to bind the receptor and deliver a signal to a neighboring cell~\cite{wright11}.
This seems a plausible idea since it provides a clear role for DeltaD as partner of DeltaC, making it necessary for synchronization.
We show that this hypothesis is compatible with data from mutants, both steady state and transient desynchronization phenotypes.
While the key component in this hypothesis is the heterodimer, we find that homodimers are also required to explain the differences in transient desynchronization of ligand mutants.
Since evidence for ligand dimers is limited, and mostly restricted to extra-embryonic assays, we then test an alternative hypothesis where monomers directly bind and activate the Notch receptor. 
We show that this model is also consistent with data, providing a role for DeltaD as a baseline signal.
These two different coupling hypotheses endow DeltaD with distinct roles.
Thus, we discuss experiments that could distinguish between the two coupling hypotheses.

The key distinction of the two scenarios is the formation of dimers that have a prominent role in binding the Notch receptors.
Dimerization of ligands has been demonstrated in-vitro using antibodies for DeltaC and DeltaD~\cite{wright11}. 
More recently, ligand dimerization has also been reported in cell culture assays~\cite{chen23}.
This work suggested that monomers bind Notch and drive trans-activation in neighbors while dimers mediate cis-inhibition.
While in-vivo colocalization has been suggested in zebrafish PSM using standard fluorescence immunohistochemistry~\cite{wright11}, 
it would be important to determine whether these dimers form as molecular species in the PSM as well, for example using a FRET assay to test ligand proximity~\cite{sun11}, or the N\&B proximity assay to test oligomerization state~\cite{digman08}.
Direct observation of in-vivo molecular dimerization would be a strong indication for a role of dimers, in favor of the dimer binding scenario.

Additionally, the two scenarios predict different dynamics for total DeltaD concentrations. 
In the dimer binding scenario, DeltaD is expected to display pronounced oscillations.
In contrast, only a weak undulation is expected in the monomer binding scenario.
Thus, a possible test between the two scenarios is to image a transgenic DeltaD protein reporter in vivo, or use antibodies, to quantify the relative amplitude of DeltaD reporter fluctuations.
Previous work using a DeltaD tagged with \textit{venus}-YFP did not report such oscillations~\cite{liao16}.
A possible caveat is that this prediction relies on the underlying assumption that ligands that bound receptors are internalized and degraded, instead of recycled.

Motivated by the distinct roles of DeltaD in both coupling hypotheses, we showed that it should be possible to provide surrogates to rescue the DeltaD mutant.
One possibility is to express the NICD independently of regulatory interactions.
This could be done by synthetic biology approaches that introduce engineered forms of the Notch receptor~\cite{morsut16, toda20, matsuda15}.
A constant level of NICD signaling or unregulated DeltaC overexpression would provide a baseline for endogenous DeltaC in the monomer binding scenario, but would fail to provide the binding partner for DeltaC in the dimer binding case.
A similar strategy would be to express exogenous DeltaC at constant levels, independent of the Her-inhibited DeltaC. 
In the monomer binding case, this additional supply of constantly expressed DeltaC would provide a baseline, while in the dimer binding case it would not be doing the job of missing DeltaD, since DeltaC homodimerization is thought to be much weaker~\cite{wright11}.
In the embryo, this could be approached by \textit{deltaC} mRNA injection, providing a pool of mRNA for translation.
%
Previous experiments using this strategy showed that injecting deltaC mRNA into either \textit{deltaC} or \textit{deltaD} heterzygotes does not improve recovery in a DAPT pulse assay, whereas injecting deltaD mRNA into the \textit{deltaD} heterzygote does~\cite{mara07}.
While these experiments revealed distinct roles for DeltaC and DeltaD ligands, the failure to rescue by deltaC mRNA is not conclusive because according to the theory rescue is non-monotonic with mRNA concentration.
A thorough test will require titrating the mRNA concentration to scan the region of possible rescue in the monomer scenario.
A limitation of this approach may be the mRNA stability, which would introduce a decreasing synthesis rate for the protein.
An alternative to circumvent this problem would be to generate a transgenic line with a constitutively expressed DeltaC, or introducing DeltaC copies with a DeltaD promoter.
The reciprocal experiment, in which a DeltaC mutant is supplemented with a DeltaD transgene, should not rescue the desynchronization phenotype.

The differential role of Notch ligands has been addressed in other contexts~\cite{boareto15, boareto16, jolly17, luna-escalante18}.
In cancer cells, a differential synthesis rate for Notch ligands Delta and Jagged was proposed to drive the formation of cell clusters that enhance metastasis~\cite{boareto16}.
In a more generic context, the roles of interacting cells as signal senders or receivers is also controlled by the amount of available Delta and Jagged, which results in the Notch-Delta-Jagged network playing a decisive role in lateral inhibition-based pattern formation~\cite{jolly17}.
Another proposal is that ligand binding efficiency of two generic ligands and competition for the same Notch binding may result in new signaling states, underlying signaling robustness and versatility~\cite{luna-escalante18}.
It would be interesting to consider these scenarios in the context of the segmentation clock in future work.

A recent study generated a library of engineered cell lines to characterize the modes of ligand receptor interaction for four different ligands and two receptors found in mammalian cells~\cite{kuintzle23}.
The study revealed a range of cis- and trans-, inhibition and activation for different concentrations.
It has also been shown that Notch cis-inhibition or -activation may play a key role in cell-fate decisions and patterning~\cite{sprinzak10, formosa-jordan14, tiedemann14}, although there is no evidence for this in the segmentation clock.

Notch signaling plays diverse roles in a wide range of developmental processes, and in disease~\cite{guruharsha12, shimojo16, aster17}. 
For example, in coordinating antiphase oscillations during neurogenesis~\cite{shimojo08, zhang18}, oscillations in fate decision of pancreatic progenitors~\cite{seymour20}, patterning of the retina ~\cite{jadhav06, mills17}, the neurogenic wavefront in the chick retina~\cite{formosa-jordan12}, tissue regeneration~\cite{gao21} and adult neurogenesis~\cite{mancini23}.
%
This remarkable versatility of Notch signaling in different contexts remains an interesting open question.
Understanding how cells use ligand multiplicity to communicate in different contexts will be relevant to medical applications such as diagnostics and therapeutics, as well as to harness this versatility for applications in gastruloid and organoid biology and tissue engineering~\cite{vandenbrink20, moris20, hofer21, arias22, yamanaka22}.

\vspace{0.5 cm}
\noindent
{\bf \normalsize Methods \\}
\\
{\bf Numerical integration.}
%
We integrate the delayed differential equations (DDE) of the model numerically using a custom code that relies on the Python library \texttt{jitcdde}~\cite{ansmann18}. 
The library implements the Shampine-Thompson algorithm using adaptive time steps~\cite{shampine01}, and casts the DDEs Python code into compiled C code for faster execution. 
%
%
The initial condition for DDEs consists of the past history, up to the largest delay into the past.
%
%
We set all variables in each cell to a constant value in the time interval $[-\tau, 0]$.
To construct steady state maps, for each cell we choose a random value from a uniform distribution in the range $[2,6]$, which initialises the system in a desynchronized state.
In desynchronization assays, we choose values in the range $[0, 0.01]$ to initialise the system in a synchronized state.
We then evolve the model for a time roughly equivalent to $100$ full cycles for steady state assays and for $30$ cycles in desynchronization assays.
For a delayed negative feedback genetic oscillator, a first order estimate of the cycle duration is $2(\tau + 1/d)$, where $\tau$ is the feedback delay and $d$ the decay rate of $h$, that we use as time scale for nondimensionalizing the model equations~\cite{lewis03, morelli07, novak08}. 
We performed all numerical assays using $N_c=100$ cells.

\noindent
\\
{\bf Wildtype and mutant parameters.}
We set wildtype parameters so that mutant conditions are consistent with experimental data.
%
%
In the embryo, DeltaC and DeltaD mutants have mean ALDs of about $5.5$ and $8$, respectively.
In the dimer scenario, we first simulate the DeltaC mutant condition, setting $\beta_C=0$, to determine its ALD value.
Since in the absence of DeltaC there is no oscillatory signal delivered to neighboring cells, the ALD of this DeltaC mutant is primarily determined by the spread of the period distribution, and is not affected by parameters $\beta_D$ and $\beta_N$.
%
%
%
%
Next, we look for $\beta_N$ and $\beta_C$ values such that DeltaD mutant condition has a mean ALD about 2.5 segments larger than that of the DeltaC mutant. 
To simulate DeltaD mutant we set $\beta_D=0$, set a $\beta_N$ value, and plot ALD as a function of $\beta_C$. 
We construct these curves for different values of $\beta_N$, and for each curve we select a wildtype $\beta_C^{\mathrm{WT}}$ value that satisfies the ALD constraint between the mutants.
We choose $\beta_N^{\mathrm{WT}}$ and $\beta_C^{\mathrm{WT}}$ to have similar values.
Finally, we choose a wildtype value $\beta_D^{\mathrm{WT}}$ so that the corresponding Kuramoto order parameter is within the range of large values, reflecting a higher level of steady state synchrony. 
With this choice, variations in $\beta_D$ of about $30\%$ do not impair wildtype synchrony levels.
To place the Notch mutant condition in the synchronization map, we decrease $\beta_N$ until we find the value that satisfies the experimental ALD of about $8$.
For the monomer scenario, we first set $\beta_D$ to be similar to the value of the dimer scenario, and chose a value for $\beta_N$ within the range that allows for steady state desynchronization of the DeltaD mutant. Then we determined the value of $\beta_C$ to match the experimental ALDs.
The existence of asymmetries between parameters for DeltaC and DeltaD synthesis and Notch binding are also consistent with experimental resynchronization assays, in which differences in ligand activities were reported~\cite{mara07}.

\noindent
\\
{\bf Synchronization.}
To quantify synchronization level of a population of genetic oscillators, we first obtain a phase $\theta_i(t)$ from the time series $h_i(t)$ using a numerical Hilbert transform~\cite{pikovsky}, as implemented in the Python package Scipy~\cite{scipy}. 
We then compute the Kuramoto order parameter $R(t)$, defined as the modulus of the average of the complex phases of individual oscillations~\cite{strogatz},
%
%
%
$R(t) e^{i \varphi(t)} = N_c^{-1} \sum_{j=1}^{N_c} e^{i \theta_j(t)}$ ,
%
%
where $\varphi(t)$ is the collective phase. 
The order parameter $R$ is close to one when all phases $\theta_i$ stay close to the collective phase, 
and drops to low values when they are spread out.
%
%
Synchrony is therefore represented by a value close to one.
To quantify steady state synchrony we take the temporal average $\bar R = \langle R(t) \rangle_t$ over a time window of about $80$ steady state cycles, ignoring the first $20$ cycles to let transients elapse.
This is the magnitude we report in synchronization maps.

\noindent
\\
{\bf ALD calculation.}
To measure the mean ALD of each mutant condition we perform $100$ desynchronizaiton assays as described above.
From each run we obtain $R(t)$ and the phase of the mean field from the Hilbert transform. 
Then, from the value of the phase at the time point where $R(t)$ crosses the threshold $R_T$ we determine how many cycles the individual realization performed before the crossing.
Since cells in the embryo go through a few cycles before the onset of somitogenesis~\cite{riedel-kruse07}, we allow for two initial cycles before starting the segment count.

\noindent
\\
{\bf DAPT assay.}
%
Notch signalling can be blocked by the drug DAPT which inhibits the proteolytic cleavage of the receptor and the consequent release of the NICD~\cite{horikawa06, riedel-kruse07}.
Above a saturating concentration $[DAPT]_\mathrm{sat} \approx 50\mu M$, no information is transmitted via the Notch pathway~\cite{herrgen10}.
%
%
%
We can simulate an assay in which a variable dose of DAPT is delivered to the embryo by including a factor in Eq.~(\ref{eq:s})
\begin{equation}
\label{eq:DAPT}
    \dot s_i = -\delta_S s_i + n_i K \left( 1-{[DAPT]}/{[DAPT]_\mathrm{sat}} \right),
\end{equation}
%
%
where $[DAPT]$ is the applied concentration of DAPT.
In desynchronziation assays, we set $[DAPT]=[DAPT]_\mathrm{sat}$, which is equivalent to setting $K=0$ in Eq.~(\ref{eq:s}).

\noindent
\\
{\bf Acknowledgments.}
We thank Bo-Kai Liao for fruitful discussions on the role of DeltaD, Melina Magalnik for illuminating conversations on experimental tests of the theory, and Margulis for entertaining contributions during discussions across timezones.
This work was supported by ANPCyT grants PICT 2017 3753 and PICT 2019 0445 awarded to LGM and FOCEM-Mercosur (COF 03/11) to IBioBA. 
%
MW was supported by a CONICET fellowship and LGM is a researcher of CONICET.
KU was supported by the JSPS KAKENHI Grant number 23K27213 and 24H00863.
ACO was supported by SNSF division III project 31003A\_176037.
LGM thanks JSPS Long Term Invitational Fellowship L23529 and the Uriu Lab at Tokyo Tech for hospitality.
%

\bibliography{DCDD.bib}

\end{document}


\cleardoublepage
\newpage
\setcounter{section}{0}
\setcounter{page}{1}

\section*{Supplementary Material}
\subsection*{Multiple Notch ligands in the synchronization \\ of the segmentation clock} %

\subsection*{by Marcos~Wappner, Koichiro~Uriu, Andrew~C.~Oates and Luis~G.~Morelli}
\vspace{0.4cm}

\section{Reaction kinetics and model formulation}
\label{sec:model}
%
%
%
{\bf Regulatory function.}
%
%
This work focuses on the roles of different components of the Notch signaling network, so we take a parsimonious approach to describe the core oscillator.
We consider a single Her protein that inhibits its own production with a time delay accounting for synthesis time.
The single variable $H(t)$ represents the concentration of this protein at time $t$.
Besides self inhibition, the synthesis rate of Her protein is  positively regulated by Notch signaling components.
The Notch intracellular domain (NICD) is released inside the cell upon binding of ligands from neighboring cells to the Notch receptor.
We denote the concentration of this signaling component by $S(t)$.

Here, we derive a function to describe the regulatory effects of $H$ and $S$ on the synthesis rate of $H$. 
We will first consider the dynamics and steady state of the promoter and the binding factors.
The promoter architecture of \textit{her} genes may be complex, including about $12$ binding sites for the Her proteins~\cite{schroter12}.
Besides, it is thought that Her proteins bind as dimers to their own promoters~\cite{schroter12}. 
It has been shown that multiple binding sites effectively increase the nonlinearity of regulatory functions~\cite{lengyel14, lengyel17, enciso13}. 
Similarly, dimerization can be effectively described in an adiabatic approximation by a Hill exponent of 2~\cite{schroter12}.
Thus, here we assume that there is only one binding site for Her proteins for simplicity, and then include the effects of multiplicity in an effective Hill exponent~\cite{lengyel14}.
We further assume that NICD binds as a single molecule to the activating complex at the Her promoter.

With these assumptions, the promoter can be in any of four states, depending on whether none, one, or both of the binding sites are occupied.
We represent the promoter state as $P_{\mu\nu}$, where $\mu$ is the state of the Her binding site and $\nu$ is the state of the NICD binding site, with $\mu, \nu = 0$ indicating an empty site and  $\mu, \nu = 1$ an occupied site.
The different promoter states are connected in the scheme of binding reactions 
\vspace{0.11cm}
\begin{center}
\begin{tikzcd}[row sep=1.5cm, column sep=1.5cm, cells={nodes={draw, outer sep=3mm, minimum height=8mm}}]
    P_{00}
        \arrow[r, harpoon, shift left, "k_{H+}"]
        \arrow[d, harpoon, shift left, "k_{S+}"] 
    & 
    P_{10}
        \arrow[l, harpoon, shift left, "k_{H-}"]
        \arrow[d, harpoon, shift left, "k_{HS+}"]
        \\
    P_{01}
        \arrow[r, harpoon, shift left, "k_{SH+}"]
        \arrow[u, harpoon, shift left, "k_{S-}"]
    &
    P_{11}
        \arrow[l, harpoon, shift left, "k_{SH-}"]
        \arrow[u, harpoon, shift left, "k_{HS-}"]
\end{tikzcd}    
\end{center}
\vspace{0.11cm}
where the $+$ and $-$ signs label binding and unbinding rates respectively. 
The values of $k_{HS\pm}$ and $k_{SH\pm}$ depend on how the two regulatory components interact. 
For instance, setting them both to zero, $k_{HS+} = k_{SH+} = 0$ describes exclusive competitive binding, where the doubly bound promoter state is not accessible~\cite{ozbudak08}. 
Alternatively, assuming the binding rates to be independent of the promoter state for both components, $k_{SH\pm}=k_{H\pm}$ and $k_{HS\pm}=k_{S\pm}$, describes a situation with no interaction at the binding sites of regulatory components, which we term dual binding model.
Further choices describe models for different ligand interactions.
Motivated by the multiple binding sites accessible to Her transcription factors, here we focus on the dual binding model.
The reactions involved in this dual binding scenario are 
\begin{align}
  \ce{P_{00} + H <=>[k_{H+}][k_{H-}] P_{10} }  \, , \qquad &\ce{P_{10} + S <=>[k_{S+}][k_{S-}] P_{11} } \, , \\
  \ce{P_{00} + S <=>[k_{S+}][k_{S-}] P_{01} }  \, , \qquad &  \ce{P_{01} + H <=>[k_{H+}][k_{H-}] P_{11} }  \, .
\end{align}

From these reactions we can write the dynamics for the four states of the promoter
\begin{align}
    \dot P_{00} &= k_{H-} P_{10} + k_{S-} P_{01} - k_{H+} H P_{00} - k_{S+} S P_{00} \, , \\
    \dot P_{10} &= k_{H+} H P_{00} + k_{S-} P_{11} - k_{H-} P_{10} - k_{S+} S P_{10} \, , \\
    \dot P_{01} &= k_{S+} S P_{00} + k_{H-} P_{11} - k_{S-} P_{01} - k_{H+} H P_{01} \, , \\
    \dot P_{11} &= k_{H+} H P_{01} + k_{S+} S P_{10} - k_{H-} P_{11} - k_{S-} P_{11}   \, , 
\end{align}
together with the conservation law
\begin{equation}
    P_{00} + P_{10} + P_{01} + P_{11} = P_{T}  	\, .
\end{equation}
We assume that promoter binding and unbinding reactions are much faster than the dynamics of $H$ and $S$.
With this assumption, the promoter state is in equilibrium for given concentrations of $H$ and $S$ and we set $\dot P_{\mu\nu} = 0 \; \forall \mu, \nu$.
We can solve the resulting algebraic equations to obtain the quasi-steady state promoter occupancy $P_{\mu\nu}(H,S)$,
\begin{align}
P_{00}(H,S) &= \frac{k_{H-} k_{S-} P_T}{\left(k_{H-}+k_{H+} H\right) \left(k_{S-}+k_{S+} S\right)}	\, , 	\\
P_{10}(H,S) &= \frac{k_{S-} k_{H+} P_T H}{\left(k_{H-}+k_{H+} H\right) \left(k_{S-}+k_{S+} S\right)}	\, , 	\\
P_{01}(H,S) &= \frac{k_{H-} k_{S+} P_T S}{\left(k_{H-}+k_{H+} H\right) \left(k_{S-}+k_{S+} S\right)}	\, , 	\\
P_{11}(H,S) &= \frac{k_{H+} k_{S+} P_T H S}{\left(k_{H-}+k_{H+} H\right) \left(k_{S-}+k_{S+} S\right)} 	\, .
\end{align}

Next, we consider how promoter occupancy determines transcription rates.
We introduce the regulatory function $f(H, S)$ that modulates the synthesis rate of $H$
\begin{equation}
    f(H, S) = \sum_{\mu,\nu=0,1} a_{\mu\nu} P_{\mu\nu}(H, S) \, , 
\end{equation}
where each promoter state $P_{\mu\nu}$ has an associated transcription rate $a_{\mu\nu}$.
Here assume that binding of Her to the promoter fully represses synthesis independently of the binding of NICD, $a_{10}=a_{11}=0$.  
We set a basal transcription rate $a_{00} = b$ and an activated state to a higher rate $a_{01} = a > b$, resulting in 
\begin{equation}
    f(H, S) = P_{T}  \frac{k_{H-} (b k_{S-} + a k_{S+} S)}{(k_{H-} + k_{H+}H)(k_{S-}+k_{S+}S)} \, .
\end{equation}
Introducing concentration scales 
\begin{equation}
H_0 \equiv \frac{k_{H-}}{k_{H+}}  \quad \mbox{and} \quad S_0 \equiv \frac{k_{S-}}{k_{S+}} \, , 
\end{equation}
we arrive at
\begin{equation}
\label{eq:linear.reg.func}
    f(H, S) = \frac{b_H}{1+\frac{H}{H_0}} \frac{1 + \frac{a_H}{b_H} \frac{S}{S_0}}{1+\frac{S}{S_0}} 	\, ,
\end{equation}
where we defined $a_H \equiv P_{T} a$ and $b_H \equiv P_{T} b$.

This expression provides the structure of the regulatory function.
As we described above, Her proteins form dimers and these dimers bind to multiple binding sites on the promoter.
Similarly, the NICD binds DNA through a complex that may also introduce additional nonlinearities.
These effects can be encompassed in effective Hill exponents $h_H$ and $h_S$, 
\begin{equation} \label{eq:regulatory.function.full}
    f(H, S) = b_H \frac{1}{1+\left(\frac{H}{H_0}\right)^{h_H}} 
    \frac{1 + \frac{a_H}{b_H} \left(\frac{S}{S_0}\right)^{h_S}}{1+\left(\frac{S}{S_0}\right)^{h_S}} \, .
\end{equation}
The regulatory function separates into factors, 
%
\begin{equation} \label{eq:regulatory.function.modules}
    f_-(H) = \frac{1}{1+\left(\frac{H}{H_0}\right)^{h_H}}  
    \quad \mbox{and} \quad
    f_+(S) = \frac{1 + \frac{a_H}{b_H} \left(\frac{S}{S_0}\right)^{h_S}}{1+\left(\frac{S}{S_0}\right)^{h_S}} ,
\end{equation}
such that
\begin{equation}
f(H, S) = b_H f_-(H) f_+(S) \, ,
\end{equation}
reflecting the independent binding assumption, where binding to one site is independent of the state of the other.

Next, to complete the formulation of the model, we describe the different reactions involved, from synthesis to degradation, ligand dimerization, and Notch binding by different components.

{\bf Synthesis.}
Her synthesis proceeds at a basal rate $b_H$, modulated by the regulatory function Eq.~(\ref{eq:regulatory.function.full}),
\begin{equation}
    \dot H_i(t) = b_H 
         \frac{1}{1+\left(\frac{H_i(t-\tau_i)}{H_0}\right)^{h_H}} 
         \frac{1+\frac{a_H}{b_H}\left(\frac{S_i(t-\tau_i)}{S_0}\right)^{h_S}}{1+\left(\frac{S_i(t-\tau_i)}{S_0}\right)^{h_S}}   + \cdots \, , 
\end{equation}
where $i=1,\ldots,N_c$ is the cell label and $\tau_i$ is an explicit synthesis delay accounting for the multiple steps in the negative feedback of $H$~\cite{mackey77, MacDonald}.
Since a Her molecule takes a time $\tau_i$ to be synthesized in cell $i$, the synthesis rate at time $t$ depends on Her and NICD concentrations at a previous time $t-\tau_i$, when synthesis was starting.
This delay is different for each cell, introducing a variability in the period of the cell population as described below.

To describe DeltaC synthesis inhibition by Her, we follow a similar strategy as above.
Considering the reaction kinetics for the binding of Her to the DeltaC promoter, 
\begin{equation}         \label{eq:synth.C}
\dot C_i(t) = b_C \frac{1}{1+\left(\frac{H_i(t-\tau_C)}{H_{0C}}\right)^{h_C}}   + \cdots \, , 
\end{equation}
where $b_C$ is the basal rate, $\tau_C$ is the DeltaC synthesis delay, $H_{0C}$ is the threshold for Her mediated inhibition of DeltaC synthesis and $h_C$ is an effective Hill exponent describing nonlinear effects in the inhibition.
Finally, we set constant synthesis rates for DeltaD and Notch, since they are not regulated by other components from the oscillator or the coupling network, resulting in constant terms
\begin{equation}
\dot D_i(t) = b_D + \cdots \qquad \mbox{and} \qquad  \dot N_i(t) = b_N + \cdots \, .
\end{equation}

{\bf Degradation.}
We assume decay of all components according to the reactions
\begin{align}
    \ce{X ->[$d_X$][] \varnothing} \, , 
    \label{eq:reaction.decay} 
\end{align}
resulting in a linear decay term
\begin{equation}
\label{eq:decay}
    \dot X_i(t) = -d_X X_i(t) + \cdots
\end{equation}
for all variables $X = H, C, D, E, F, G, N, S$, 
where $d_X$ is the decay rate for the variable $X$.

{\bf Ligands dimerization.}
Notch ligands DeltaC and DeltaD can form an heterodimer and both homodimers, described in the reactions
\begin{align}
    \ce{C_i + D_i <=>[k_{CD}^+][k_{CD}^-] C\colon \!\!D_i} 		\label{eq:reaction.dimerization.E} \, , \\
    \ce{2C_i <=>[k_{CC}^+][k_{CC}^-] C\colon \!\!C_i} 			\label{eq:reaction.dimerization.F} \, , \\
    \ce{2D_i <=>[k_{DD}^+][k_{DD}^-] D\colon \!\!D_i} 			\label{eq:reaction.dimerization.G} \, .
\end{align}
We introduce the notation for dimers concentrations, $\mbox{[DeltaC:DeltaD]} \equiv E$, $\mbox{[DeltaC:DeltaC]} \equiv F$, $\mbox{[DeltaD:DeltaD]} \equiv G$, and relabel the corresponding rates as $k_E^{\pm} \equiv k_{CD}^{\pm}$, $k_F^{\pm} \equiv k_{CC}^{\pm}$, and $k_G^{\pm} \equiv k_{DD}^{\pm}$.
Each reaction contributes terms to the components involved.
For example, Eq.~\eqref{eq:reaction.dimerization.F} will contribute terms to the equations for the DeltaC monomer and the DeltaC:DeltaC homodimer,
\begin{align}
    \dot C_i(t)	&= - 2k_F^+ C_i^2 + 2k_F^- F_i + \cdots  \, , \\ 
    \dot F_i(t)	&= + k_F^+ C_i^2 - k_F^- F_i +   \cdots  \, .
\end{align}

{\bf Notch binding.} 
Ligands from a neighboring cell, in the form of monomers or dimers, may bind to a Notch receptor.
Once Notch is bound by a ligand, its intracellular domain is released into the cell and the complex formed by the remaining extracellular domain and the bound ligand is internalized into the other cell. 
Here we assume that the ligand is degraded after the interaction with a receptor.
Since the receptor is cleaved in the process, we describe this as an irreversible reaction,
\begin{equation}
\label{eq:reaction.activation.X} 
    \ce{X_i + N_j ->[k_{NX}] S_j} \qquad , \qquad \mbox{X = C, D, E, F, G} \, , 
\end{equation}
where $k_{NX}$ is the binding rate between Notch and a ligand $X$, and $i, j$ label the interacting cells.

{\bf Average neighbor coupling.}
These Notch binding reactions give rise to the coupling terms in the dynamics of $X$
\begin{equation}
    \dot X_i(t) = - \sum_{j\in\mathcal{V}_i} {k_{NX} X_i N_j r_{ij}} + \cdots \, , 
\end{equation}
where the sum is over neighbors $\mathcal{V}_i$ of cell $i$.
The contribution from each neighbor $j$ is weighted by the fraction $r_{ij}$ of the total surface of cell $i$ in contact with cell $j$,
\begin{equation}
r_{ij} = \frac{A_{ij}}{A_T} \, , 
\end{equation}
where $A_{ij}$ is the contact area between the two neighbors and $A_T$ is the total area of a cell, assumed to be the same for all cells.
Assuming that the total cell $i$ area is shared equally with all $|\mathcal{V}_i|$ neighbors, $A_{ij}=A_T/|\mathcal{V}_i|$, and
\begin{equation}
r_{ij} = \frac{1}{|\mathcal{V}_i|}.
\end{equation}
With this assumption, cell $i$ is coupled to the average of Notch concentrations in neighboring cells,
\begin{equation} \label{eq:avex}
\dot X_i(t) = -k_{NX} X_i \frac{1}{|\mathcal{V}_i|} \sum_{j\in\mathcal{V}_i} {N_j} + \cdots  \, .
\end{equation}
Similar average couplings arise in the equations for $N$ and $S$,
\begin{align} 
\dot N_i(t) &= -k_{NX} N_i \frac{1}{|\mathcal{V}_i|} \sum_{j\in\mathcal{V}_i} {X_j} + \cdots  \, ,   \label{eq:aven} \\
\dot S_i(t) &= +k_{NX} N_i \frac{1}{|\mathcal{V}_i|} \sum_{j\in\mathcal{V}_i} {X_j} + \cdots  \, .   \label{eq:aves}
\end{align}

{\bf Mean field.}
In this work we focus on the tailbud, a posterior region of the segmentation clock characterized by synchronized oscillations in a population of very mobile cells~\cite{uriu17}.
This cell mobility causes a continuous neighbor exchange, and it has been shown that the resulting dynamics can be effectively described as an all-to-all coupling, or mean field~\cite{uriu13}.
Thus, we write the average terms in Eqs.~(\ref{eq:avex})-(\ref{eq:aves}) for the ligands and the Notch receptor as mean field variables
\begin{equation} \label{eq:mf}
\bar X = \frac{1}{N_c}\sum_{i=1}^{N_c} X_i \qquad , \qquad X = C, D, E, F, G, N .
\end{equation}
With these definitions, we can write the gain and loss terms that reactions in Eq.~\eqref{eq:reaction.activation.X} contribute to different components.
For example, the heterodimer E participates in the contributions
\begin{align}
    \dot S_i(t) &= + k_{NE} \bar E N_i  + \cdots \, , \\
    \dot E_i(t) &= - k_{NE} E_i \bar N  + \cdots \, , \\ 
    \dot N_i(t) &= - k_{NE} \bar E N_i  + \cdots \, .
\end{align}

{\bf Full model.}
Putting all contributions together, we have eight coupled delayed differential equations for each cell $i$, 
\begin{widetext}  
	\begin{align}
	    \dot H_i &= -d_H H_i + b_H 
	         \frac{1}{1+\left(\frac{H_i(t-\tau_i)}{H_0}\right)^{h_H}}
	         \frac{1+\frac{a_H}{b_H}\left(\frac{S_i(t-\tau_i)}{S_0}\right)^{h_S}}
	            {1+\left(\frac{S_i(t-\tau_i)}{S_0}\right)^{h_S}}
	         \label{eq:dde.dim.H}\\
	    \dot C_i &= -d_C C_i + b_C \frac{1}{1+\left(\frac{H_i(t-\tau_C)}{H_{0C}}\right)^{h_C}}
	         - k_E^+ C_i D_i + k_E^-E_i - 2k_F^+ C_i^2 + 2k_F^- F_i - k_{CN} C_i \bar N 
	         \label{eq:dde.dim.C}\\
	    \dot D_i &= -d_D D_i + b_D + k_E^-E_i - k_E^+ C_i D_i 
	         - 2k_G^+ D_i^2 + 2k_G^- G_i- k_{DN} D_i \bar N
	         \label{eq:dde.dim.D}\\
	    \dot E_i &= -d_E E_i - k_E^- E_i + k_E^+ C_i D_i - k_{EN} E_i \bar N
	         \label{eq:dde.dim.E}\\
	    \dot F_i &= -d_F F_i - k_F^- F_i + k_F^+ C_i^2 - k_{FN} F_i \bar N
	         \label{eq:dde.dim.F}\\
	    \dot G_i &= -d_G G_i - k_G^- G_i + k_G^+ D_i^2 - k_{GN} G_i \bar N
	         \label{eq:dde.dim.G}\\
	    \dot N_i &= -d_N N_i + b_N - k_{CN} \bar C N_i - k_{DN} \bar D N_i - k_{EN} \bar E N_i
	        - k_{FN} \bar F N_i - k_{GN} \bar G N_i
	         \label{eq:dde.dim.N}\\
	    \dot S_i &= -d_S S_i + k_{CN} \bar C N_i + k_{DN} \bar D N_i + k_{EN} \bar E N_i
	        + k_{FN} \bar F N_i + k_{GN} \bar G N_i
	         \label{eq:dde.dim.S}
	\end{align}
\end{widetext}  
where we omit the time dependence for notational simplicity except in delayed contributions.
This full model is a superset of all the special cases considered in the main text, including all possible dimers and Notch binding species.
The scenarios considered in the main text can be obtained by an adequate choice of parameters.
For instance, choosing the binding rates $k_{CN} = k_{DN} = 0$ gives us the dimer binding scenario of the main text, while choosing all dimerization constants $k_{X}^{\pm}=0$  and the corresponding Notch coupling constants $k_{NX}=0$ for $X = E, F, G$ results in the monomer binding scenario.

{\bf Period variability.}
The autonomous period of the core oscillator $i$ depends primarily on the value of the synthesis delay $\tau_i$ in the negative feedback, and the half life of the protein $d_H^{-1}$, with second order corrections depending mainly on the synthesis rate $b_H$~\cite{lewis03}.
To introduce variability in the autonomous periods, we sample the delays $\tau_i$ of individual oscillators from a normal distribution, with mean $\tau_H$ and variance $\sigma_\tau$.

\section{Dimensionless formulation}
\label{sec:adim}
Next, we introduce the dimensionless formulation of the model that we use in the main text, which provides a description that is independent of timescales and concentration scales.
The equations (\ref{eq:dde.dim.H})-(\ref{eq:dde.dim.S}) have units of concentration over time.
We choose the concentration scale $H_0$ and the timescale $d_H^{-1}$, and render the equations dimensionless multiplying all by the same factor $(d_H H_0)^{-1}$.
Introducing dimensionless variables 
\begin{equation}
x_i \equiv \frac{X_i}{H_0}  \quad , \quad  X \ = H, C, D, E, F, G, N, S \, ,
\end{equation}
and dimensionless time
\begin{equation}
\tilde t \equiv t d_H \, ,
\end{equation}
the time derivatives transform as
%
%
%
\begin{equation}
\frac{\dot X}{d_H H_0} = \frac{dx}{d\tilde t} \equiv x'  \, ,
\end{equation}
where we defined the prime notation for the derivative with respect to $\tilde t$.
The resulting dimensionless formulation is
\begin{widetext}  
	\begin{align}
	        h_i' &= -h_i + \beta_h 
	        \frac{1} {\left[ 1 + h_i^{\eta_H}(\tilde t- \tilde \tau_i) \right]}
		\frac{ 1 +\alpha \left( \sigma s_i(\tilde t- \tilde \tau_i) \right) ^{\eta_S} }
		{ \left[ 1 + \left( \sigma s_i(\tilde t- \tilde \tau_i) \right) ^{\eta_S} \right] }
	       \label{eq:dde.adim.h} \\
	    c_i' &= - \delta_c c_i + \beta_c \frac{1}{1 + \left(\gamma h_i(\tilde t- \tilde \tau_C)\right)^{\eta_C} } 
	        + \lambda_E^- e_i - \lambda_E^+ c_i d_i + \lambda_F^- f_i - \lambda_F^+ c_i^2 - \kappa_C c_i \bar n
	       \label{eq:dde.adim.c} \\
	    d_i' &= - \delta_d d_i + \beta_d 
	        + \lambda_E^- e_i - \lambda_E^+ c_i d_i + \lambda_G^- g_i - \lambda_G^+ d_i^2 - \kappa_D d_i \bar n
	       \label{eq:dde.adim.d} \\
	    e_i' &= - \delta_e e_i - \lambda_E^- e_i + \lambda_E^+ c_i d_i - \kappa_E e_i \bar n 
	       \label{eq:dde.adim.e} \\
	    f_i' &= - \delta_f f_i - \lambda_F^- f_i + \lambda_F^+ c_i^2 - \kappa_F f_i \bar n 
	       \label{eq:dde.adim.f} \\
	    g_i' &= - \delta_G g_i - \lambda_G^- g_i + \lambda_G^+ d_i ^2 - \kappa_G g_i \bar n 
	       \label{eq:dde.adim.g} \\
	    n_i' &= - \delta_n n_i + \beta_n - n_i \sum_{c, d, e, f, g} \kappa_X \bar x
	       \label{eq:dde.adim.n} \\
	    s_i' &= - \delta_s s_i + n_i \sum_{x = c, d, e, f, g}\kappa_X \bar x 
	    \label{eq:dde.adim.s} 
	\end{align}
\end{widetext}  
with dimensionless parameters
\begin{align*}
    \delta_X \equiv \frac{d_X}{d_H}   \quad &, \quad
    \beta_X = \frac{b_X}{d_H H_0}     \quad , \quad
    \kappa_X = \frac{k_{XN}H_0}{d_H}  \, ,
    \\
    \sigma = \frac{H_0}{S_0}          \quad &, \quad 
    \gamma \equiv \frac{H_0}{H_{0C}}  \quad , \quad
    \alpha = \frac{a_H}{b_H}          \, ,
\end{align*}
\begin{align*}
    \lambda_E^+ \equiv \frac{H_0 k_E^+}{d_H}  \quad &, \quad \lambda_E^- \equiv \frac{k_E^-}{d_H}      \, ,\\
    \lambda_F^+ \equiv \frac{2H_0 k_F^+}{d_H} \quad &, \quad \lambda_F^- \equiv \frac{2 k_F^-}{d_H}    \, ,\\
    \lambda_G^+ \equiv \frac{2H_0 k_G^+}{d_H} \quad &, \quad \lambda_G^- \equiv \frac{2 k_G^-}{d_H}    \, .
\end{align*}
and renaming $h_X \rightarrow \eta_X$, $X=H, S, C$.
For notational simplicity and readability, in the main text we drop the tildes from dimensionless time variables $\tilde t \rightarrow t$ and $\tilde \tau_X \rightarrow \tau_X$. 
We also drop the prime notation and use a dot to denote dimensionless time derivatives.

\section{Parameter values}
\label{sec:}
%
Next, we parametrize the model to describe both wildtype and mutant conditions. 
Experimental observations of the zebrafish segmentation clock, together with theoretical considerations, set constraints on parameter values. 
To parametrize the core oscillator we follow~\cite{lewis03}, comparing the mRNA and protein model
\begin{align}
\dot p(t) &= a m(t-\tau_p) - b p(t)  \, ,  \label{eq:lewisp}  \\
\dot m(t) &= \frac{k}{1+\left(\frac{p(t-\tau_m)}{p_0} \right)^n} - c m(t)   \, , \label{eq:lewism}
\end{align}
to the protein only oscillator from Eq.~(\ref{eq:dde.adim.h})
\begin{equation} \label{eq:coreH}
\dot H(t) = -d_H H(t) + \frac{b_H}{1+\left(\frac{H(t-\tau_H)}{H_0} \right)^{\eta_H} } \, .
\end{equation}
Parameter values estimated for the mRNA and protein oscillator are~\cite{lewis03}: 
$a=4.5$ proteins per mRNA molecule,
$c=0.23$ molecules per minute,
$b=0.286$ molecules per minute estimated from experiments~\cite{ay14},
$k=33$ mRNA molecules per diploid cell per minute, and
$p_0=40$ molecules. 
Synthesis delays in the model were later estimated from experiments~\cite{hanisch13}, showing that her1 and her7 have very similar transcription times $\tau_{m1} \approx 10$ min and $\tau_{m7} \approx 9$ min, dominating over the protein translation times $\tau_{p1} \approx 2.8$ min and $\tau_{p7} \approx 1.7$ min.

Introducing a quasi-steady state approximation for the mRNA, $\dot m \approx 0$, we obtain
\begin{equation} \label{eq:mapprox}
m(t) \approx \frac{k}{c} \frac{1}{1+\left(\frac{p(t-\tau_m)}{p_0} \right)^n} 	\, ,
\end{equation}
and replacing $m(t-\tau_p)$ into (\ref{eq:lewisp}),
\begin{equation} \label{eq:lewispp}
\dot p(t) = -b p(t) + \frac{ak}{c} \frac{1}{1+\left(\frac{p(t-\tau_m-\tau_p)}{p_0} \right)^{n} } \, .
\end{equation}
Comparing Eqs.~(\ref{eq:coreH}) and (\ref{eq:lewispp}) we see that to obtain similar dynamics we should set $d_H \approx b$, $b_H \approx ak/c$, $\tau_H \approx \tau_m + \tau_p$, $\eta_H \approx n$ and $H_0 \approx p_0$.
Since this comparison relies on the quasi-steady state approximation, a strict assignment of these values may produce differences in amplitude and period.
In addition, the synthesis delay $\tau_H$ could be estimated differently from comparing the estimated period for both models,
\begin{align}
T_{\mbox{\footnotesize Lewis}}    &\approx 2 (\tau_p +\tau_m + 1/b + 1/c)  \label{eq:perlewis}  \, , \\
T_H                 &\approx 2 (\tau_H + 1/d_H)   \, . \label{eq:perH}
\end{align}
For these periods to match, we would require $\tau_H \approx \tau_m + \tau_p + 1/c$.

For the protein only model, parameter values from~\cite{lewis03} imply a Her protein degradation rate $d_H = 0.286$ min$^{-1}$, setting the timescale, and $H_0 = p_0 = 40$ molecules, corresponding to a concentration scale $[H_0] \approx 0.3$ nM for tailbud cells of 7.6 $\mu$m in diameter~\cite{uriu17}.
%
%
To obtain oscillations of similar amplitude to~\cite{lewis03}, we set $\beta_H = 28$ in the dimensionless model, corresponding to $b_H \approx 320$ proteins per minute. 
We set the dimensionless delay to $\tau_h = 4.2$, corresponding to a synthesis delay within a range of values obtained from averaging Her1 and Her7 data, $\tau_H = 14.7$.
%
%
Hill exponents have an effective value that encompasses the effect of multiple macromolecular interactions in the regulation, such as dimerization~\cite{schroter12}, binding of complexes and multiple binding sites at the promoter~\cite{lengyel14, lengyel17}. 
We expect exponent values to be larger than 2, so we set the conservative estimate $\eta = 2.5$.
%
%
We will use these values for the core oscillator and determine the rest of parameters from further estimations and explorations.
\begin{table}[t!]
\begin{center}
\begin{tabular}{@{}|c|c|c|l|@{}}
\toprule
\hline
parameter      & dimers            & monomers & description \\ \midrule
\hline\hline
$\delta_C$     & 1                 & 1     & DC degradation rate \\
\hline
$\delta_D$     & 1                 & 1     & DD degradation rate \\
\hline
$\delta_E$     & 1                 & 1     & DC:DD degradation rate \\
\hline
$\delta_F$     & 1                 & 1     & DC:DC degradation rate \\
\hline
$\delta_G$     & 1                 & 1     & DD:DD degradation rate \\
\hline
$\delta_N$     & 1                 & 1     & Notch degradation rate \\
\hline
$\delta_S$     & 1                 & 1     & NICD degradation rate \\
\hline\hline
\addlinespace
$\beta_H$      & 28                & 28    & Her synthesis rate  \\
\hline
$\beta_C$      & 26                & 39    & DC synthesis rate  \\
\hline
$\beta_D$      & 17                & 20    & DD synthesis rate  \\
\hline
$\beta_N$      & 24                & 30    & Notch synthesis rate  \\
\hline\hline
\addlinespace
$\lambda_{E+}$ & 1                 & –     & DC:DD dimerization rate\\
\hline
$\lambda_{E-}$ & 0.1               & –     & DC:DD dissociation rate\\
\hline
$\lambda_{F+}$ & 0.1               & –     & DC:DC dimerization rate\\
\hline
$\lambda_{F-}$ & 0.1               & –     & DC:DC dissociation rate\\
\hline
$\lambda_{G+}$ & 0.1               & –     & DD:DD dimerization rate\\
\hline
$\lambda_{G-}$ & 0.1               & –     & DD:DD dissociation rate\\
\hline\hline
\addlinespace
$\kappa_C$     & –                 & 0.02  & DC binding rate to Notch \\
\hline
$\kappa_D$     & –                 & 0.01  & DD binding rate to Notch \\
\hline
$\kappa_E$     & 0.2               & –     & DC:DD binding rate to Notch \\
\hline
$\kappa_F$     & 0.02              & –     & DC:DC binding rate to Notch \\
\hline
$\kappa_G$     & 0.009             & –     & DD:DD binding rate to Notch \\
\hline\hline
\addlinespace
$\sigma$       & 0.1               & 0.1   & Signal activation threshold \\
\hline
$\gamma$       & 1                 & 1     & DC inhibition threshold\\
\hline
$\eta_H$       & 2.5               & 7     & Her self-inhibition H.e.\\
\hline
$\eta_S$       & 2.5               & 7     & Her activation by signal H.e.\\
\hline
$\eta_C$       & 2.5               & 7     & DC inhibition by Her H.e.\\
\hline
$\alpha$       & 10                & 10    & Fold change in synthesis rate\\
\hline
$\tau_H$       & 4.2               & 4.2   & Mean Her synthesis delay \\
\hline
$\sigma_\tau/\tau_H$& 0.03         & 0.03  & C.V. of Her synthesis delay \\
\hline
$\tau_C/\tau_H$& 1.7               & 1.7   & Relative DC synthesis delay \\
\hline
\bottomrule
\end{tabular}
\caption{Parameter table used in synchronization maps in the main text. 
DC and DD stand for DeltaC and DeltaD. 
H.e. stands for Hill exponent.
C.V. is coefficient of variation.}
\end{center}
\label{tab:parameters}
\end{table}

The rest of the parameters concern signaling components, such as the ligands DeltaC and DeltaD, their dimerization, binding to Notch receptors and signal production and action on the oscillator.
We set some of these parameters below and explore the effects of varying others.

Next we consider signal reception parameters in Eq.~(\ref{eq:dde.adim.h}).
The dimensionless threshold for signal driven synthesis activation $\sigma$ controls how much signal is necessary to affect the core oscillator. We chose a value $\sigma = 0.1$, which corresponds to a weaker action than the inhibitor. 
For the signal to activate Her protein synthesis above basal levels we require a coupling strength $\alpha > 1$. 
Through exploration, we settled for a value $\alpha = 10$, which increases Her collective oscillation amplitude by a factor of $\sim 2$. 
The corresponding Hill exponent $\eta_S$ was set equal to $\eta_H$ for simplicity.

In the regulation of DeltaC synthesis by Her protein Eq.~(\ref{eq:dde.adim.c}), we set the dimensionless 
threshold for synthesis repression $\gamma  = 1$, assuming for simplicity that the same concentration of Her protein is required to inhibit both Her and DeltaC synthesis.
Experimental observations in zebrafish showed that impairing Notch signaling results in longer collective period~\cite{herrgen10}.
For reduced coupling to lengthen the period, effective coupling delays should be slightly below the period value~\cite{schuster89, morelli09}.
With this motivation we set the synthesis delay of DeltaC to $\tau_C = 1.7 \tau_H$. 
The corresponding Hill exponent $\eta_C$ was set equal to $\eta_H$ for simplicity.

Given the lack of specific data for each component, we set all decay rates in the model to the same value $d_X = d_H$ with $X \in \{C, D, E, F, G, N, S \}$, for simplicity.
So $\delta_X = 1$ for all $X$ in the dimensionless model.
Motivated by experimental observations, we set the values of dimerization and binding rates to Notch of the two homodimers about an order or magnitude lower than those of the heterodimer~\cite{wright11}. 
Scans of these parameters revealed that dimer association rates $\lambda_X^+$ have maximal effective values: beyond some level, the size of the synchronization region in $\beta_C$ vs. $\beta_D$ space stops changing significantly. 
Dimer dissociation rates $\lambda_X^-$ were found to have small effects on the size and shape of the synchronization regions as long as they where within an order of magnitude of the corresponding $\lambda_X^+$, and were set to $\lambda_X^-=0.1$.

The remaining parameters are the synthesis rates of coupling components and the binding rates of different components to Notch receptors.
We vary synthesis rates $\beta_C$, $\beta_D$ and $\beta_N$ to construct synchronization maps in terms of these.
Additionally, we vary the values of $\kappa_X$ to explore how these maps change.

\newpage
\bibliography{DCDD.bib}

\newpage
\onecolumngrid

\setcounter{figure}{0}
\renewcommand{\figurename}{{\bf Fig.}}
\renewcommand{\thefigure}{{\bf S\arabic{figure}}}

\section{Supplementary Figures}
\begin{figure}[h]
\centering
\includegraphics[]{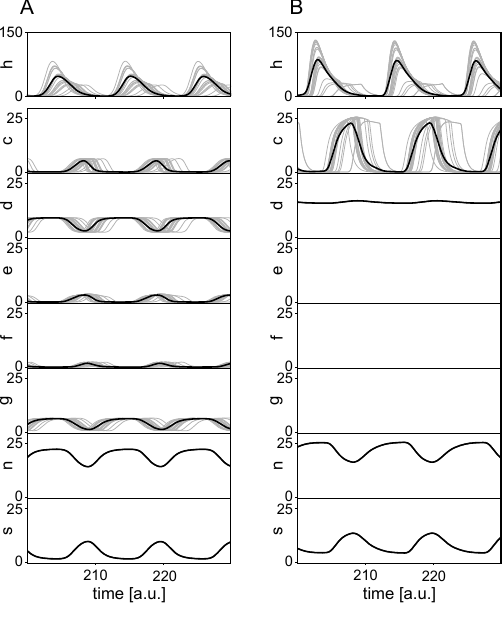}
\caption{Steady state individual oscillations $x_i(t)$ (grey lines, 25 out of 100 are displayed) and 
mean field $\bar x(t)$ (black line), for all variables $x=h, c, d, e, f, g, n, s$, in the (A) dimer scenario and (B) monomer scenario. 
$e(t)$, $f(t)$ and $g(t)$ are not defined in the monomer model, hence panels are empty. 
Vertical scales are the same except for $h(t)$.
Parameters as in Table~\ref{tab:parameters}.}
\label{fig:traces}
\end{figure}
\begin{figure}[h]
\centering
\includegraphics[]{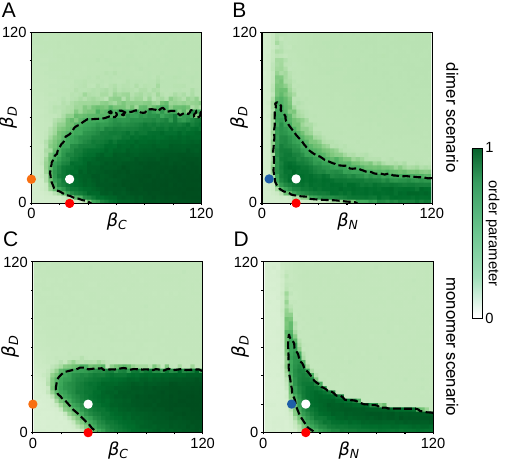}
\caption{Extended synchronization maps for wider ranges of the syntheses rates as indicated, for (A, B) the dimer scenario shown in Fig. 2 of the main text and (C, D) the monomer scenario shown in Fig. 5 of the main text. Colored dots indicate the parameter sets used in the desynchronization assays for each model.
Dashed line and number of realizations as in Fig.~2C, D of the main text. 
}
\label{fig:extended}
\end{figure}
\begin{figure}[h]
\centering
\includegraphics[]{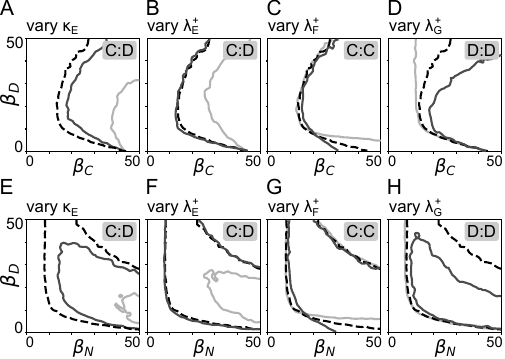}
\caption{Changes to the sync region boundary of the map 
(A-D) in Fig. 2C of the main text and 
(E-H) in Fig. 2D of the main text, 
as a function of different parameters of the model.
(A, E) $\kappa_E=0.02$ (light), $0.066$ (dark), $0.2$ (dashed black).
(B, F) $\lambda_E^+ = 0.05$ (light), $1$ (dashed black), $3.33$ (dark).
(C, G) $\lambda_F^+ = 0.006$ (light), $0.1$ (dashed black), $2$ (dark).
(D, H) $\lambda_G^+ = 0.006$ (light), $0.1$ (dashed black), $2$ (dark).
Dashed line and number of realizations as in Fig.~2E-F of the main text. 
}
\label{fig:dimer_thresholds}
\end{figure}
\begin{figure}[h]
\centering
\includegraphics[]{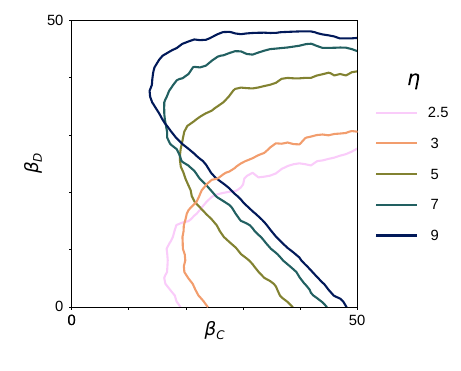}
\caption{Changes to the sync region boundary of the monomer binding scenario map in Fig. 5B of the main text as a function of the Hill exponent $\eta$, which controls the nonlinearity of the regulatory function. 
$\eta=2.5$ corresponds to Fig.~5B of the main text and $\eta=7$ to Fig.~5C. 
Sync region boundaries and number of realizations as in Fig.~2E-F of the main text. 
}
\label{fig:monomer_thresholds}
\end{figure}%
\begin{figure}[h]
\centering
\includegraphics[]{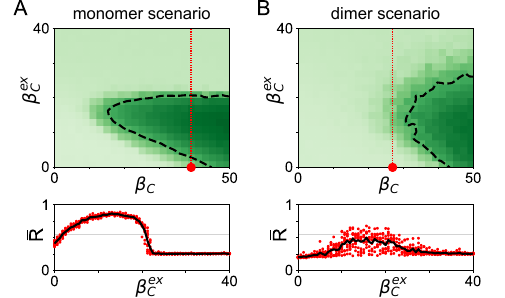}
\caption{Exogenous DeltaC expression assay in DeltaD mutant predicts distinct outcomes for monomer and dimer scenarios.
We included an additional synthesis term in Eq.~(\ref{eq:dde.adim.c}), $c_i' = ... + \beta_C^{ex}$, where $\beta_C^{ex}$ is an exogenous constant synthesis rate of DeltaC.
Top: Steady state order parameter $\bar R$ in terms of endogenous and exogenous DeltaC synthesis rates $\beta_C$ and $\beta_C^{ex}$ for the (A) monomer and (B) dimer binding scenarios.
Red dot is the DeltaD mutant condition for $\beta_C^{ex} = 0$ and vertical red dotted line indicates the cut plotted in bottom panels.
Dashed line, color scale and number of realizations as in Fig.2C, D of the main text. 
Bottom: Steady state order parameter $\bar R$ as a function of $\beta_C^{ex}$ along the red dotted line in top panels for (A) monomer and (B) dimer binding scenarios.
Red dots are 10 independent realizations and black line is the average.
Horizontal gray line indicates the threshold $R_T$.
Parameters as in corresponding scenario, except for $\beta_D=0$, Table~\ref{tab:parameters}.
}
\label{fig:rescue_sup}
\end{figure}